\input harvmac
\input psfig
\newcount\figno
\figno=0
\def\fig#1#2#3{
\par\begingroup\parindent=0pt\leftskip=1cm\rightskip=1cm\parindent=0pt
\global\advance\figno by 1
\midinsert
\epsfxsize=#3
\centerline{\epsfbox{#2}}
\vskip 12pt
{\bf Fig. \the\figno:} #1\par
\endinsert\endgroup\par
}
\def\figlabel#1{\xdef#1{\the\figno}}
\def\encadremath#1{\vbox{\hrule\hbox{\vrule\kern8pt\vbox{\kern8pt
\hbox{$\displaystyle #1$}\kern8pt}
\kern8pt\vrule}\hrule}}
\def\underarrow#1{\vbox{\ialign{##\crcr$\hfil\displaystyle
 {#1}\hfil$\crcr\noalign{\kern1pt\nointerlineskip}$\longrightarrow$\crcr}}}
%
\overfullrule=0pt

%

\def\bar{\overline}
\def\Z{{\bf Z}}

\def\R{{\bf R}}

\font\zfont = cmss10 
\font\litfont = cmr6

\def\bigone{\hbox{1\kern -.23em {\rm l}}}
\def\ZZ{\hbox{\zfont Z\kern-.4emZ}}
\def\half{{\litfont {1 \over 2}}}

\Title{hep-th/0006071}
{\vbox{\centerline{}
\bigskip
\centerline{Noncommutative Tachyons And String Field Theory }}}
\smallskip
\centerline{Edward Witten}
\smallskip
\centerline{\it Dept. of Physics, Cal Tech, Pasadena, CA}
\smallskip\centerline{and}
\smallskip
\centerline{\it CIT-USC Center For
Theoretical Physics, USC, Los Angeles CA}

\medskip

\noindent
It has been shown recently that by turning on a large noncommutativity
parameter, the description of tachyon condensation in string theory
can be drastically simplified.  We reconsider these issues from
the standpoint of string field theory, showing that, from this point of view,
the key fact is that in the limit of a large $B$-field,
the string field algebra factors as the product of an algebra
that acts on the string center of mass only and an algebra that acts
on all other degrees of freedom carried by the string.
\Date{June, 2000}

 \nref\strom{R. Gopakumar, S. Minwalla,
and A. Strominger, ``Noncommutative Solitons,'' hep-th/0003160.}%
\nref\dmr{K. Dasgupta, S. Mukhi, and G. Rajesh,
``Noncommutative Tachyons,'' hep-th/005006}
\nref\hklm{J. A. Harvey, P. Kraus, F. Larsen, and E. Martinec,
``$D$-Branes And Strings As Noncommutative Solitons,'' hep-th/0005031,}
Recently, solitons in scalar field
theories on noncommutative spacetimes with very large noncommutativity
parameter $\theta$ have been constructed
in a strikingly simple way \strom.  This insight has been applied 
\refs{\dmr,\hklm} to string
theory with strong noncommutativity  to test the
predictions about tachyon condensation and brane annihilation \ref\sen{A.
Sen, ``Tachyon Condensation On The Brane Antibrane System,'' hep-th/9805170,
``$SO(32)$ Spinors Of Type I And Other Solitons On Brane-Antibrane
Pair,'' hep-th/9808141, ``Descent Relations Among Bosonic $D$-Branes,'' hep-th/9902105,
Int. J. Mod. Phys. {\bf A14} (1999) 4061, ``Universality Of The Tachyon
Potential,'' JHEP {\bf 9912:027} (1999).}.

The purpose of the present paper is to 
examine these issues in the context of open string field theory,
as formulated in the language of noncommutative geometry
\ref\witten{E. Witten, ``Noncommutative Geometry and String Field Theory,''
Nucl. Phys. {\bf B268} (1986) 253.}.  String field theory has been used to explore tachyon  condensation
\nref\kostsam{V. A. Kostelecky and S. Samuel, ``On A Nonperturbative Vacuum
For The Open Bosonic String,'' Nucl. Phys. {\bf B336} (1990) 263.}%
\nref\othersen{A. Sen and B. Zwiebach, ``Tachyon Condensation In
String Field Theory,'' hep-th/9912249, JHEP {\bf 0003:002} (2000).}%
\nref\tayl{W. Taylor, ``$D$-Brane Effective Field Theory From String Field
Theory,'' hep-th/0001201; N. Moeller and W. Taylor, ``Level Truncation
And The Tachyon In Open Bosonic String Field Theory,'' hep-th/0002237.}
\nref\othertach{N. Berkovits, ``The Tachyon Potential In Open 
Neveu-Schwarz String Field Theory,'' hep-th/0001084;
N. Berkovits, A. Sen, and B. Zwiebach,  ``Tachyon
Condensation In Superstring Field Theory,'' hep-th/0002211.}%
\nref\gopt{R. de Mello Koch, A. Jevicki, M. Mihailescu, and R. Tatar,
``Lumps And $p$-Branes In Open String Field Theory,'' hep-th/0003031.}%
\nref\sr{P. De Smet and J. Raeymaekers, ``Level Four Approximation To The
Tachyon Potential In Superstring Field Theory,'' hep-th/0003220.}%
\nref\iqbal{A. Iqbal and A. Naqvi, ``Tachyon Condensation On A Non-BPS
$D$-Brane,'' hep-th/0004015.}%
\refs{\kostsam - \iqbal}, generally at $\theta=0$.
As one might expect
\nref\cds{A. Connes, M. R. Douglas, and A. S. Schwarz,  ``Noncommutative
Geometry And Matrix Theory: Compactification On Tori,'' hep-th/9711162,
JHEP {\bf 9802:003} (1998).}%
\nref\seiwit{N. Seiberg and E. Witten, hep-th/9908142, JHEP
{\bf 9909:032} (1999).}%
\refs{\cds,\seiwit}, there is considerable simplification for large
$\theta$, and so we will in this paper consider in string field theory
language the solutions studied  in \refs{\strom - \hklm}.  We will carry
out this discussion in flat $\R^{26}$ or $\R^{10}$.

\bigskip\noindent{\it Factorization Of The Algebra}

\def\A{{\cal A}}
\def\V{{\cal V}}
In string field theory, the starting point is an associative algebra
$\A$ that is built by multiplying string fields.  The naive idea for
constructing such an algebra is to start with the association of
open string states with vertex operators $\V$.  An open string vertex
operator is of course inserted at the boundary of the open string,
at some proper time $\tau$ along the boundary.  Naively speaking,
to multiply string states we would like to just multiply the corresponding
vertex operators.  The trouble is that the product $\V\cdot \V'$
of open string vertex operators at the same point on the boundary
is not well-defined, because of the familiar short distance singularities
of products of local quantum field operators.  

We do have an operator product expansion (OPE)
\eqn\ucin{\V(\tau)\V'(\tau')\rightarrow\sum_kc_k|\tau-\tau'|^{-a_k}
\V_k(\tau')~~{\rm for}~\tau\to\tau'.}
The coefficients $c_k$ in this expansion depend on whether $\tau>\tau'$ or
$\tau<\tau'$; this is the origin of noncommutativity.
There does not seem to be any elegant way to eliminate the $\tau$ dependence
 and  extract an associative algebra $\A$ from the operator
product expansion.  In open string field theory, this is done by making
rather special choices of local parameters for insertions of vertex
operators; the construction is perhaps most naturally described in 
terms of gluing of open string states \witten.   
At any rate,  in this paper, we will only need
properties of the algebra $\A$ that follow in a very general way from
the properties of the operator product algebra.  Technical details in
the definition of $\A$ will not be important.

The operator product expansion conserves the ``ghost number'' of the
vertex operators, and hence $\A$ is graded by ghost number.  The classical
string field $A$ is a ghost number one element of $\A$.  The worldsheet
BRST operator $Q$ is a ghost number one derivation of the algebra
(that is, $Q(A*A')=QA*A'+(-1)^AA*QA'$), and the  equation of
motion of bosonic open string field theory is
\eqn\imoco{QA+A*A=0.}
A similar construction can be made for open superstrings, but has been
argued to have technical difficulties associated with meeting of
picture-changing operators on the worldsheet \ref\whog{C. Wendt, ``Scattering
Amplitudes And Contact Interactions In Witten's Superstring Field Theory,''
Nucl. Phys. {\bf B314} (1989) 209; J. Greensite and F. R. Klinkhamer,
``Superstring Amplitudes And Contact Interactions,'' Nucl. Phys. {\bf B304}
(1988) 108.}.
A modification has been proposed to circumvent this difficulty
\ref\berk{N. Berkovits, ``Super-Poincar\'e Invariant Superstring Field
Theory,'' Nuc. Phys. {\bf B450} (1995) 90, hep-th/9503099,
``A New Approach To Superstring Field Theory,'' hep-th/9912121, in
proceedings of the 32$^{nd}$ Symposium Ahrenshoop on the Theory
of Elementary Particles, Fortschritte der Physik {\bf 48} (2000) 31; N. Berkovits
and C. T. Echevarria, ``Four-Point Amplitude From Open Superstring Field
Theory,'' hep-th/9912120.}.  The effect of this is to replace \imoco\ by a 
nonpolynomial equation which, for our purposes in the present paper,
can be treated in precisely the same way.

Now, the OPE algebra of open string vertex operators has a subalgebra  
$\A_0$
in which one does not use the string center of mass coordinate.
Thus, $\A_0$ contains vertex operators 
that depend in an arbitrary fashion on the ghosts $b$ and $c$
and the derivatives of the spacetime coordinates $X^i$, $i=1,\dots,26$,
all taken at zero spacetime momentum.
$\A_0$ contains, for example, $b\partial c(\partial X^1)^{22}
\partial^3X^2$, but not $b\partial c(\partial X^1)^{22}
\partial^3X^2e^{ip\cdot X}$ with $p\not= 0$.  Operators of zero momentum
are closed under OPE's, and so $\A_0$ is a subalgebra of $\A$.

One may ask whether there is a complementary subalgebra $\A_1$ generated
{\it only} by the $e^{ip\cdot X}$ and without $\partial X$, $\partial^2X$,
etc.  Normally, the answer to this question is ``no,'' since even if one
merely makes a classical Taylor series expansion, the operator products
of exponentials involve also the derivatives of $X$: 
\eqn\polgo{e^{ip\cdot X}(\tau)e^{iq\cdot X}(\tau')
\to e^{i(p+q)\cdot X}(\tau')+i(\tau-\tau')
p\cdot \partial X e^{i(p+q)\cdot X}(\tau')+\dots.}

However, there is a  limit in which one actually can  factorize $\A$ in 
terms of commuting subalgebras  as $\A=\A_0\otimes
\A_1$, where  as suggested above
$\A_0$ consists of vertex operators of zero momentum
and $\A_1$ is generated by operators $e^{ip\cdot X}$.
This is the limit in which the NS two-form
field $B$ is constant and large \refs{\cds,\seiwit}.  In fact,
we assume that $B$ is of maximal rank in 10- or 26-dimensional Euclidean
space.  

To see this factorization, we first recall
\nref\frts{E. Fradkin and A. Tseytlin, ``Nonlinear Electrodynamics
From Quantized Strings,'' Phys. Lett. {\bf B163} (1985) 123.}%
\nref\ctny{C. G. Callan, A. Tseytlin, C. R. Nappi, and S. Yost,
``String Loop Corrections To Beta Functions,'' Nucl. Phys. {\bf B288} (1987)
525;  A. Abouelsaood, C. G. Callan, C. R. Nappi, and S. A. Yost,
``Open Strings In Background Gauge Fields,'' Nucl. Phys. {\bf B280} (1987)
599.}%
the form of the worldsheet propagator in the presence of a $B$-field.
We take the string world-sheet to consist of the upper half plane.
The propagator between boundary points $\tau,\tau'$ on the real axis
is then \refs{\frts,\ctny}
\eqn\kolgo{\langle X^i(\tau)X^j(\tau')=-\alpha'\left({1\over
g+2\pi \alpha' B}\right)^{ij}_S\ln(\tau-\tau')^2+i\pi\alpha'\left({1\over
g+2\pi \alpha' B}\right)^{ij}_A\epsilon(\tau-\tau').}
Here $g$ is the closed string metric and $(~)_S$, $(~)_A$ denote the symmetric
and antisymmetric part of a matrix.

Now we take the limit $B\to\infty$ with $g,\alpha'$ fixed.  To be definite,
take
$B=tB_0$ with $t\to\infty$.  (This is the same limit as in \seiwit,
but parametrized differently.)
If we set 
\eqn\ploog{X^i=Y^i/\sqrt t,}
 then the propagator becomes
\eqn\polgo{\langle Y^i(\tau)Y^j(\tau')\rangle ={1\over t(2\pi)^2\alpha'}\left(\theta^2\right)^{ij}\ln(\tau-\tau')^2+{i\over 2} \theta^{ij}\epsilon(\tau-\tau'),}
where $\theta=1/B_0$.  

Now for $t\to\infty$, the $e^{iq\cdot Y}$ do generate a closed algebra
as the $\ln(\tau-\tau')^2$ term does 
not contribute.  (We will make this more explicit momentarily.) This is the center of mass algebra
$\A_1$.  What about $\A_0$, the algebra that {\it doesn't} contain
the center of mass momentum?  As $\partial Y(\tau)\cdot \partial Y(\tau')
\sim t^{-1}/(\tau-\tau')^2$, we see that if we use $\sqrt t\, \partial^nY$
as the algebra generators of $\A_0$, then the structure constants are
independent of $t$.   A typical element of $\A_0$ is then
\eqn\gurf{b\partial c \cdot \sqrt t \,\partial^{n_1}Y^{a_1}\cdot
\sqrt t \,\partial^{n_2}Y^{a_2}\cdot \sqrt t\, \partial^{n_3}Y^{a_3},}
with a factor of $\sqrt t$ for each $\partial^nY$.  

So in this limit, we have two algebras $\A_0$ and $\A_1$.  They commute
and the full algebra is $\A=\A_0\otimes \A_1$.  To verify that $\A_0$ and
$\A_1$ commute, we simply have to observe that
\eqn\commo{\sqrt t\, 
\partial^nY (\tau)e^{iq\cdot Y}(\tau')\sim {1\over \sqrt t}
(\tau-\tau')^{-n}
e^{iq\cdot Y},}
since the only term in the propagator that contributes is the logarithmic
term, proportional to $t^{-1}$.  The right hand side vanishes for $t\to
\infty$.  Finally, to verify that $\A_1$ is closed under OPE's, we note
that when we expand
\eqn\mimo{e^{ip\cdot Y}(\tau)e^{iq\cdot Y}(\tau')
\sim e^{-\half i\theta_{ij}p^iq^j}e^{i(p+q)\cdot Y}(\tau')(1+i(\tau-\tau')
p\cdot \partial Y
+\dots),}
the corrections proportional to $\partial Y$ can be dropped since the factor
of $\partial Y$ is not accompanied by a factor of $\sqrt t$.
The OPE algebra $\A_1$ thus reduces to 
\eqn\mimom{e^{ip\cdot Y}(\tau)e^{iq\cdot Y}(\tau')
\sim e^{-\half i\theta_{ij}p^iq^j}e^{i(p+q)\cdot Y}(\tau')}
which is the algebra of functions on noncommutative $\R^{10}$ or
$\R^{26}$; as explained in \seiwit,
the usual complications of the OPE disappear, because the dimensions vanish and the right hand side has no dependence on $\tau-\tau'$.

So far we have assumed that the $B$-field has maximal rank.
If $B$ has less than maximal rank, we modify the factorization
so that $\A_1$ is generated by $e^{ip\cdot X}$ where $p$ is tangent to the
noncommutative directions and $\A_0$
is generated by all other operators.  We still get a factorization
$\A=\A_0\otimes \A_1$ in terms of commuting subalgebras.
In this more general case, an operator in $\A_0$ may carry
momentum, but in commutative directions only,
while $\A_1$, on the other hand, will be the algebra of functions in the
noncommutative directions.
$\A_1$ is a down-to-earth algebra that can be described concretely
in finite-dimensional terms, while $\A_0$ contains all of the
mysterious stringy complications.

By a similar scaling, the BRST operator Q acts on $\A_0$ and commutes
with $\A_1$.  So the string field equation $0=QA+A*A$ makes
sense for $A\in \A_0$.

\bigskip\noindent{\it Tachyon Condensation}

For bosonic strings, there are at least
two important solutions known with $A\in \A_0$.
One of them is $A=0$ and describes the ordinary open string vacuum.
The other, which will here be called $A_0$, was first explored
numerically in \kostsam\ and is now understood to describe tachyon
condensation to ``nothing,'' that is, to a state with only closed strings.
This means, in particular, that the corresponding $A_0$-$A_0$ open strings,
with boundary conditions at both ends determined by the classical
open string solution $A_0$, have no physical excitations.

Now more generally, we could introduce $2\times 2$ Chan-Paton
factors and start with two $D25$-branes.  The solution
\eqn\junko{A=\left(\matrix{ 0 & 0 \cr 0 & A_0 \cr}\right)}
describes annihilation to a state with just one $D25$-brane.  There
are now several kinds of open strings:

(1) The $0$-$0$ open strings describe physical excitations of the surviving
$D25$-brane.

(2) The $0$-$A_0$ and $A_0$-$A_0$ open strings are expected to have no
physical excitations.

Since the solution $A_0$ lies in $\A_0$, which (after suitably rescaling
the coordinates) is completely independent
of $B$, the solution $A_0$ is completely insensitive to the $B$-field.
Now, let us specialize to the limit of large $B$ and
invoke the idea of \refs{\strom}.
Let $\rho\in\A_1$ be any projection operator, that is any element with 
\eqn\kikop{\rho^2=\rho.}
Then as $[Q,\rho]=0$, we see that $A=A_0\otimes \rho$ obeys
$0=QA+A*A$ if $A_0$ does.  Equivalently, since \kikop\ implies
that $(1-\rho)^2=(1-\rho)$, we can solve the equation with
\eqn\nikop{A=A_0\otimes (1-\rho).}

\def\H{{\cal H}}
Now, following \strom, represent $\A_1$ as the algebra of operators
on a Hilbert space ${\cal H}$.  The endpoint of a string has a Chan-Paton
label that takes
values in ${\cal H}$.   Take $\rho$ to be the projector onto
a finite-dimensional subspace of $\H$, say a subspace $V$ of dimension $n$.
Write $\H=V\oplus W$ where $W$ is the orthocomplement of $V$;
so $\rho|_V=1$ and $\rho|_W=0$.  In expanding around the solution
$A=A_0\otimes (1-\rho)$, we get several kinds of open strings:

(1)$'$  The $V$-$V$ open strings, that is strings each of whose endpoints
are labeled by vectors in $V$, are 
governed by the usual equations of open string theory except that
the effective open string algebra for these strings is $\A_0\otimes M_n$,
where $M_n$ is the algebra of $n\times n$ complex matrices acting on $V$.
Hence the momentum of these strings is always
tangent to the commutative directions in
spacetime, and there are effective $n\times n$ Chan-Paton factors.
These modes describe the physical excitations of $n$ $D(25-2p)$-branes,
where $2p$ is the number of noncommutative directions. 

(2)$'$ The $V$-$W$ and $W$-$W$ open strings are governed by the same
equations that describe the $0$-$A_0$ and $A_0$-$A_0$ open strings
discussed above; so if the usual conjectures about tachyon condensation are
true, then these open strings have no physical excitations.

Thus, this solution describes annihilation of a $D25$-brane to
a collection of $n$ parallel $D(25-2p)$-branes, for arbitrary $p$ and $n$.

\bigskip\noindent{\it Type IIA}

At the very formal level of our discussion, we can consider tachyon
condensation for unstable $D9$-branes of Type IIA in much the same way. 
In factorizing $\A=\A_0\otimes \A_1$,
we include the superconformal
ghosts and worldsheet fermions
 in $\A_0$; $\A_1$ is as in the case of the bosonic string the algebra
of functions in the noncommuting directions of spacetime.
There is, conjecturally,
 still a solution $A_0$ that describes tachyon condensation,
and a more general solution $A=A_0\otimes (1-\rho)$ which, if $\rho$ is the
projection operator to an $n$-dimensional subspace, describes annihilation
to a system of $n$ $D(9-2p)$-branes.  As noted in \refs{\dmr,\hklm}, 
a further subtlety arises  because for open string excitations
of a Type IIA $D9$-brane, there is a $\Z_2$
symmetry that changes the sign of the tachyon field.  Let $A_0'$ be the
conjugate solution with opposite tachyon field.  
The solutions $A_0$ and $A_0'$ describe tachyon condensation to two different
closed string vacua that differ by the sign of the tachyon field;
in fact, they differ by one unit of the Ramond-Ramond zero-form
$G_0/2\pi$ \ref\horava{P. Horava,
``Type IIA $D$-Branes, $K$ Theory, and Matrix Theory,'' hep-th/9812135,
Adv. Theor. Math. Phys. {\bf 2} (1999) 1373.}. 

If $\rho_1,\rho_2$ obey
$\rho_1^2=\rho_1$, $\rho_2^2=\rho_2$, $\rho_1\rho_2=\rho_2\rho_1=0$, we can
make a more general solution with $A=A_0\otimes\rho_1+A_0'\otimes \rho_2$.
A special case of this with $\rho_1=\rho$, $\rho_2=1-\rho_1$ is
\eqn\jinglo{A=A_0\otimes(1-\rho)+A_1\otimes \rho.}
Let us consider $\rho$ to be the projector onto all quantum states
that are supported within a large region $\Omega$ in the noncommutative
phase space.  This is only a rough, semiclassical description of $\rho$,
but it should be good if $\Omega$ is large enough.  \jinglo\ describes
tachyon condensation to a state in which the tachyon field has one
sign outside of $\Omega$ and another sign inside $\Omega$; because
the two states differ by one unit of $G_0/2\pi$, there is a supersymmetric $D8$-brane wrapped on the boundary of $\Omega$.  
As noted in \refs{\dmr, \hklm}, it is perplexing that in this approximation
the tension of the $D8$-brane appears to be zero. 

To explore this puzzle in a little more detail (but without claiming
to resolve it), consider the case
of two noncommutative directions with coordinates $x,y$ obeying
$[x,y]=-i\theta$.  Suppose that $\Omega$ is a disc and that we
want $\rho$ to be a projector onto an $n$-dimensional subspace.
Then the area of $\Omega$ should be $A=2\pi\theta n$, so its radius
is $r=\sqrt{2\theta n}$.  $\rho$ is approximately 1 deep inside $\Omega$
and approximately 0 far from $\Omega$; the scale of variation of $\rho$
is approximately the same as the width in space of the outermost quantum
state onto which $\rho$ projects.  (If we try to make $\rho$ vary more slowly
than that, there will be states on which it is not equal approximately to
either 0 or 1.)   That outermost state
fills a cylindrical shell near the boundary of $\Omega$
of area $2\pi\theta$; the radial thickness of the shell is thus
$\Delta r = \theta/r=\sqrt{\theta/2n}$.  The validity of our
description rests on neglecting the logarithmic term in \polgo, which
is proportional to $\theta^2/t\alpha'$;
this term can be considered small if the scale of variation of the solution
is large compared to $\theta/\sqrt{t\alpha'}$.  The condition we
need is thus $\Delta r>>\theta/\sqrt{t\alpha'}$ or 
\eqn\uvvu{{A\over t}<<\alpha'.}
Since $A/t$ is the area of $\Omega$ in the original coordinates
$X$, before the rescaling \ploog, this means that the solution \jinglo\
is actually only valid if the area of $\Omega$ in string units is much
less than one.

\bigskip\noindent{\it Type IIB}

Now what can we say about tachyon condensation for Type IIB superstrings?
For Type IIB, we could first of all consider a $D9$- or  $D\bar 9$-brane
with $A=0$.  The corresponding boundary conditions for open strings
we will call $0$ and $\bar 0$, respectively.

The combined $D9$-$D\bar 9$ system is believed to admit a somewhat more
interesting solution.   First of all, to describe a $D9$-$D\bar 9$ system,
we use $2\times 2$ Chan-Paton matrices, but with the opposite GSO projection
for the off-diagonal terms.  Thus the string field takes the form
\eqn\polko{A=\left(\matrix{ B & T \cr \bar T & B'\cr}\right),}
where $B$ and $B'$ have the usual GSO projections and $T$ and $\bar T$
have the opposite ones; thus $B$ and $B'$ describe gauge fields as
well as stringy excitations, while $T$ and $\bar T$ have a tachyon at the 
lowest level.  There is a symmetry
\eqn\imno{T\to e^{-i\theta}T,~~\bar T \to e^{i\theta}\bar T.}
It is believed that there exists a solution 
that describes tachyon condensation to the closed string vacuum.
It has been explored numerically \refs{\othertach,\sr,\iqbal} but
is not known in closed form; we merely write it as
\eqn\jimno{A_0=\left(\matrix{\alpha & \beta \cr \bar\beta & \gamma}\right)}
Because of the symmetry \imno, it can be generalized to
\eqn\himno{A_\theta=
\left(\matrix{\alpha & e^{-i\theta}\beta \cr e^{i\theta}\bar\beta & \gamma}\right).}

As in the discussion of the bosonic string, we can add extra uncondensed
$D9$'s and $D\bar 9$'s and mix the two solutions.  Thus, in a larger
space, we can consider the string field
\eqn\onono{A=\left(\matrix{0 & 0 \cr
                           0 & A_\theta\cr}\right),}
where the upper left block describes excitations of a $D9$ or $D\bar 9$.
This field certainly obeys the equations of string field theory, and describes
partial annihilation of a system of $D9$'s and $D\bar 9$'s, leaving a
single brane.  The open strings in expanding around this
solution can be classified as $0$-$0$ open strings, which describe
ordinary open string excitations, and $0$-$\theta$, $\theta$-$0$,
and $\theta$-$\theta$ open strings, all of which describe no physical modes
at all if the usual hypotheses about tachyon condensation are correct.

In the large $B$ limit, the string field algebra factors 
as $\A=\A_0\otimes \A_1$ just as before. We want to generalize the solution $A_\theta\in \A_0$ to include
the $\A_1$ factor, by a suitable generalization of the previous ansatz
$A=A_0\otimes \rho$.

For this, we let $\sigma$ be an element of $\A_1$ and $\bar \sigma$ its complex
conjugate (or hermitian adjoint in a Hilbert space representation).
We take
\eqn\jugglo{A=\left(\matrix{\alpha\otimes \bar\sigma\sigma & \beta\otimes 
\bar\sigma\cr
  \bar \beta\otimes  \sigma & \gamma\otimes  \sigma \bar\sigma\cr}
\right).}  To
obey $QA+A*A=0$ given that $A_0$ obeys this equation,
\foot{And similarly for any other equation, like the one in \berk,
that is constructed by multiplication of string fields as well as operations
like $Q$ that commute with $\A_1$.} the properties we need are
\eqn\mimomo{\sigma\bar \sigma \sigma = \sigma, ~~\bar \sigma\sigma\bar \sigma=  \bar \sigma.}

If $\sigma$ is invertible, these equations say that $\sigma$ is unitary.  On
an eigenspace with $\sigma=e^{i\theta}$, we get 
\eqn\onond{A=A_\theta=\left(\matrix{ \alpha &  e^{-i\theta}\beta \cr
                                     e^{i\theta}\beta & \gamma\cr}\right).}
This solution describes tachyon condensation to a state with no
physical excitations.

The fun comes when $\sigma$ is not invertible.  
Let $V$ be the kernel of $\sigma$, and $W$ the kernel of $\bar\sigma$
(or equivalently the cokernel of $\sigma$).  
Let $n$ and $m$ be the dimensions of
$V$ and $W$, and $M_n$, $M_m$ the algebras of matrices acting on $V$ and
$W$ respectively; we suppose that $n$ and $m$ are finite.  The equations
\mimomo\ mean in general that $\sigma$ is a unitary isomorphism between the 
orthocomplement of $V$ and the orthocomplement of $W$.

If $2p$ is the number of noncommutative directions, then
string states whose endpoints are labeled by vectors in $V$ describe
the excitations of $n$ $D(9-2p)$-branes, and those whose endpoints
are labeled by vectors in $W$ describe the excitations of $m$ $D(\bar{9-2p})$-
branes.  The $V$-$W$ open strings are equivalent to
conventional $D(9-2p)$-$D(\bar{9-2p})$ open strings.  Other open
string excitations of this system are governed by the same equations
as the $0$-$\theta$
and $\theta$-$\theta$ excitations of the the solution \onono, and describe
no physical excitations at all, if the conventional hypotheses are correct.
This solution thus describes tachyon condensation down to a system
with $n$ $D(9-2p)$-branes and $m$ $D(\bar{9-2p})$-branes. 
The net $D(9-2p)$-brane charge is $n-m$, which is the same as the index of
the operator $\sigma$.\foot{This solution has been described in a more
detailed setting in section 4 of \hklm.  Note in eqn. (4.8) of that
paper an operator of index 1.}

To describe explicitly a solution of \mimomo\ with nonzero index, suppose
that there are two noncommutative directions with coordinates $x,y$, with 
\eqn\jullo{[x,y]=-i\theta, ~~\theta>0.}
We introduce the creation and annihilation operators
\eqn\polo{a={x-iy\over \sqrt{2\theta}},  ~~\bar a={x+iy\over \sqrt{2\theta}},
~~  [a,\bar a]=1.}
$a$ and $\bar a$ are represented on a Hilbert space ${\cal H}$ that contains
a vector $|0\rangle$ with $a|0\rangle=0$.  The kernel of
$a$ is generated by $|0\rangle$, and $\bar a$ has no kernel.
We let
\eqn\pullo{\eqalign{\sigma & = {1\over \sqrt{\bar a a+1}}a\cr
                     \bar\sigma & = \bar a{1\over\sqrt{\bar a a + 1}}.\cr}}
Clearly, the kernel of $\sigma$ is generated by $|0\rangle$, and
the kernel of $\bar\sigma$ is empty; so the index of $\sigma$ is 1. A short computation shows that
$\sigma\bar\sigma=1$, which implies \mimomo.
From \polo, we have
\eqn\kolop{\sigma= {1\over \sqrt{x^2+y^2+\theta}}(x-iy).}
If $\sigma$ is treated as a classical function of $x$ and $y$, then
for $x,y\to\infty$ we have $|\sigma|=1$.  Ignoring the commutator $[x,y]$
is valid in describing the behavior near infinity.
We can thus regard $\sigma$ near
infinity as a $U(1)$-valued function on a circle; as such, its winding
number is $-1$.  Thus, in this particular case, the index equals minus
the winding number.  This relation is a special case of the
general Atiyah-Singer index theorem.
Since the index and the winding number are both topological invariants,
the relation between the index and the winding number
 can be proved, in this particular problem, by computing the index for one operator of
every possible index, for example $\sigma=(1/\sqrt{ a^n\bar a^n}) a^n$
for positive index or $\sigma = \bar a^n(1/\sqrt{a^n\bar a^n})$ for negative
index.  We leave details to the reader.  
The identification of the $D$-brane charge with the winding number of
the tachyon field was the original proposal in \sen.

\bigskip\noindent{\it Electric Flux Tubes?}

In this paper, we have formulated in a slightly more abstract
language many solutions that were described in \refs{\dmr,\hklm}.
A notable exception is the solution describing fundamental strings
as electric flux tubes that is proposed in section 5 of \hklm.
This solution appears to depend on properties of the string theory
effective action that are more specific than the general features that
have been exploited in the present paper.

\bigskip
This paper is based on a talk presented at the Lennyfest, Stanford
University, May 20-1, 2000.
The work was supported in part by NSF Grant PHY-9513835 and the
Caltech Discovery Fund.
\listrefs
\end